\documentclass[twoside,twocolumn,9pt]{article}
\usepackage{extsizes}
\usepackage[super,sort&compress,comma]{natbib} 
\usepackage[version=3]{mhchem}
\usepackage[left=1.5cm, right=1.5cm, top=1.785cm, bottom=2.0cm]{geometry}
\usepackage{balance}
\usepackage{times,mathptmx}
\usepackage{sectsty}
\usepackage{graphicx} 
\usepackage{lastpage}
\usepackage[format=plain,justification=justified,singlelinecheck=false,font={stretch=1.125,small,sf},labelfont=bf,labelsep=space]{caption}
\usepackage{float}
\usepackage{fancyhdr}
\usepackage{fnpos}
\usepackage[english]{babel}
\usepackage{array}
\usepackage{droidsans}
\usepackage{charter}
\usepackage[T1]{fontenc}
\usepackage[usenames,dvipsnames]{xcolor}
\usepackage{setspace}
\usepackage[compact]{titlesec}
\usepackage{hyperref}

\usepackage{graphics}
\usepackage{epsfig}
\usepackage{amsmath}
\usepackage{amssymb}
\usepackage{color}

\usepackage{epstopdf}

\definecolor{cream}{RGB}{222,217,201}

\begin{document}

\pagestyle{fancy}
\thispagestyle{plain}
\fancypagestyle{plain}{

\fancyhead[C]{\includegraphics[width=18.5cm]{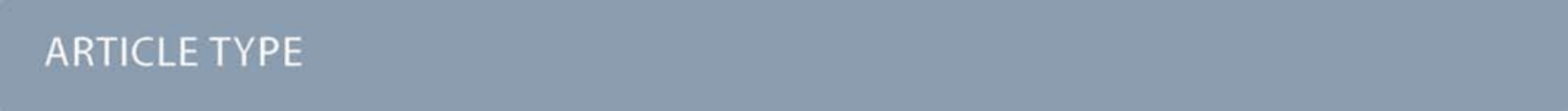}}
\fancyhead[L]{\hspace{0cm}\vspace{1.5cm}\includegraphics[height=30pt]{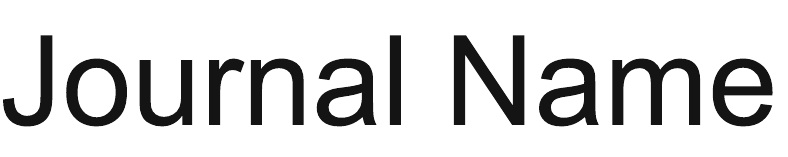}}
\fancyhead[R]{\hspace{0cm}\vspace{1.7cm}\includegraphics[height=55pt]{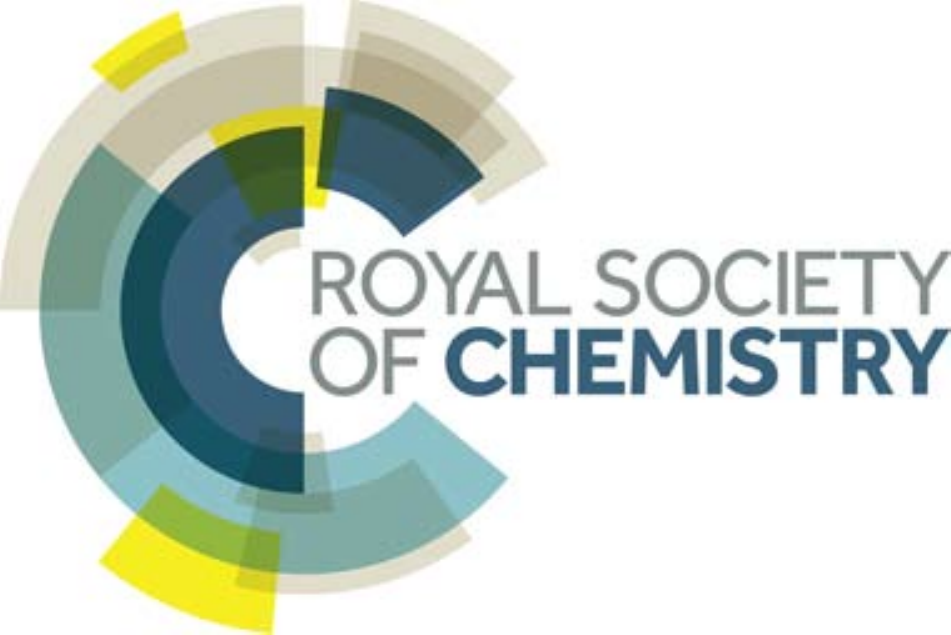}}
\renewcommand{\headrulewidth}{0pt}
}

\makeFNbottom
\makeatletter
\renewcommand\LARGE{\@setfontsize\LARGE{15pt}{17}}
\renewcommand\Large{\@setfontsize\Large{12pt}{14}}
\renewcommand\large{\@setfontsize\large{10pt}{12}}
\renewcommand\footnotesize{\@setfontsize\footnotesize{7pt}{10}}
\makeatother

\newcommand\brian[1]{{\color{green}#1}}

\renewcommand{\thefootnote}{\fnsymbol{footnote}}
\renewcommand\footnoterule{\vspace*{1pt}%
\color{cream}\hrule width 3.5in height 0.4pt \color{black}\vspace*{5pt}} 
\setcounter{secnumdepth}{5}

\makeatletter 
\renewcommand\@biblabel[1]{#1}            
\renewcommand\@makefntext[1]%
{\noindent\makebox[0pt][r]{\@thefnmark\,}#1}
\makeatother 
\renewcommand{\figurename}{\small{Fig.}~}
\sectionfont{\sffamily\Large}
\subsectionfont{\normalsize}
\subsubsectionfont{\bf}
\setstretch{1.125} 
\setlength{\skip\footins}{0.8cm}
\setlength{\footnotesep}{0.25cm}
\setlength{\jot}{10pt}
\titlespacing*{\section}{0pt}{4pt}{4pt}
\titlespacing*{\subsection}{0pt}{15pt}{1pt}

\fancyfoot{}
\fancyfoot[LO,RE]{\vspace{-7.1pt}\includegraphics[height=9pt]{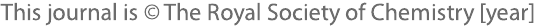}}
\fancyfoot[CO]{\vspace{-7.1pt}\hspace{13.2cm}\includegraphics{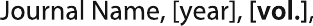}}
\fancyfoot[CE]{\vspace{-7.2pt}\hspace{-14.2cm}\includegraphics{head_foot/RF}}
\fancyfoot[RO]{\footnotesize{\sffamily{1--\pageref{LastPage} ~\textbar  \hspace{2pt}\thepage}}}
\fancyfoot[LE]{\footnotesize{\sffamily{\thepage~\textbar\hspace{3.45cm} 1--\pageref{LastPage}}}}
\fancyhead{}
\renewcommand{\headrulewidth}{0pt} 
\renewcommand{\footrulewidth}{0pt}
\setlength{\arrayrulewidth}{1pt}
\setlength{\columnsep}{6.5mm}
\setlength\bibsep{1pt}

\makeatletter 
\newlength{\figrulesep} 
\setlength{\figrulesep}{0.5\textfloatsep} 

\newcommand{\topfigrule}{\vspace*{-1pt}%
\noindent{\color{cream}\rule[-\figrulesep]{\columnwidth}{1.5pt}} }

\newcommand{\botfigrule}{\vspace*{-2pt}%
\noindent{\color{cream}\rule[\figrulesep]{\columnwidth}{1.5pt}} }

\newcommand{\dblfigrule}{\vspace*{-1pt}%
\noindent{\color{cream}\rule[-\figrulesep]{\textwidth}{1.5pt}} }

\makeatother

\twocolumn[
  \begin{@twocolumnfalse}
\vspace{3cm}
\sffamily
\begin{tabular}{m{4.5cm} p{13.5cm} }

\includegraphics{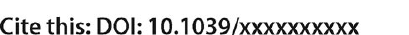} & \noindent\LARGE{\textbf{Softening and Yielding of Soft Glassy Materials}} \\
\vspace{0.3cm} & \vspace{0.3cm} \\

 & \noindent\large{
 Simon Dagois-Bohy,\textit{$^{a,b}$} 
 Ell\'ak Somfai,\textit{$^{c}$} 
 Brian P. Tighe,$^{\ast}$\textit{$^{d}$} 
 and Martin van Hecke\textit{$^{a,e}$}} \\

\includegraphics{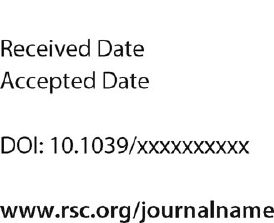} & \noindent\normalsize{
Solids deform and fluids flow, but soft glassy materials, such as emulsions, foams, suspensions, and pastes, exhibit an intricate mix of solid and liquid-like behavior. While much progress has been made to understand their elastic (small strain) and flow (infinite strain) properties, such
understanding is lacking for the softening and yielding phenomena that connect these asymptotic
regimes. Here we present a comprehensive framework for softening and yielding of soft glassy materials, based on extensive numerical simulations of oscillatory rheological tests, and show that two distinct scenarios unfold depending on the material's packing density. For dense systems, there is a single, pressure-independent strain where the elastic modulus drops and the particle motion becomes diffusive. In contrast, for weakly jammed systems, a two-step process arises: at an intermediate softening strain, the elastic and loss moduli both drop down and then reach a new plateau value, whereas the particle motion becomes diffusive at the distinctly larger yield strain. We show that softening is associated with an extensive number of microscopic contact changes leading to a non-analytic rheological signature. Moreover, the scaling of the  softening strain with pressure suggest the existence of a novel pressure scale above which softening and yielding coincide, and we verify the existence of this crossover scale numerically. Our findings thus evidence the existence of two distinct classes of soft glassy materials -- jamming dominated and dense -- and show how these can be distinguished by their  rheological fingerprint.
} \\
\end{tabular}
 \end{@twocolumnfalse} \vspace{0.6cm}
  ]

\renewcommand*\rmdefault{bch}\normalfont\upshape
\rmfamily
\vspace{-1cm}


\footnotetext{\textit{$^{a}$Huygens-Kamerlingh Onnes Lab, Leiden University, P.O.~Box 9504, 2300 RA Leiden, The Netherlands}}
\footnotetext{\textit{$^{b}$~Laboratoire de M\'ecanique des Fluides et d'Acoustique, UMR CNRS 5509 Universit\'e de Lyon, INSA de Lyon 36, Avenue Guy de Collongue 69134 \'Ecully cedex, France }}
\footnotetext{\textit{$^{c}$~Institute for Solid State Physics and Optics, Wigner Research Center for Physics, Hungarian Academy of Sciences, P.O.~Box 49, H-1525 Budapest, Hungary }}
\footnotetext{\textit{$^{d}$~Delft University of Technology, Process \& Energy Laboratory, Leeghwaterstraat 39, 2628 CB Delft, The Netherlands; email: b.p.tighe@tudelft.nl }}
\footnotetext{\textit{$^{e}$~AMOLF, Science Park 104, 1098 XG Amsterdam, The Netherlands }}

%


The rheology of soft glassy materials is an intricate mixture of elastic, viscous and plastic behaviors. Oscillatory rheology is an ideal tool to characterize these materials, as variation of the driving frequency and driving amplitude allows one to quantify the relative importance of elastic, viscous and plastic contributions \cite{barnes}. Depending on the driving conditions, these materials exhibit both a solid-like and liquid-like regime: for vanishingly small strain amplitude, the material's response is linear and can be characterized by an elastic and a loss modulus - in solids the former exceeds the latter at low frequencies and the material behaves elastically. In contrast, for sufficiently large strain amplitudes the materials yields and flows. Near the yield strain, the elastic modulus drops and, in many cases, the loss modulus peaks. What are the rheological scenarios that connect  the small and large strain regimes? What are the microscopic signatures associated with increased driving strains? Which aspects of these scenarios depend on material properties, and which aspects are universal? Here we use extensive numerical simulations of particle-based models of non-Brownian amorphous solids to disentangle  how the pre- and post-yielding regimes are connected.
We evidence that there are two qualitatively distinct scenarios for the strain dependent response of soft glassy materials, depending on the rigidity of the {\em jammed state} of the material at rest.

The jamming transition has in recent years been shown to play an important organizing role for the response of soft repulsive sphere packings, which are an effective model for
foams, emulsions, suspensions and granular media \cite{durian95,ohern03,katgert10b,majmudar07}.
First, there is overwhelming evidence that the distance to the critical jamming point, as measured by e.g.~the confining pressure $P$, is the key control parameter governing a packing's quasistatic, linear elastic response \cite{ohern03,wyart05,ellenbroek06,boschan16,baumgarten17}, and more recent work has extended these findings to the linear viscoelastic response at finite frequency \cite{tighe11,boschan16,boschan17,baumgarten17b}. Second, the distance to jamming has also been found to play a key role in organizing the steady state rheology of a wide range of soft materials \cite{olsson07,heussinger09,tighe10c,nordstrom10,ikeda13}. As  linear response and steady state rheology are connected to the limits of zero and infinite strain amplitude in oscillatory shear, the distance to jamming can be expected to play a crucial role at finite strains as well. Moreover, several characteristic scales that limit the range of linear behavior all scale with the distance to jamming: when the pressure is lowered towards the critical point, the material becomes dominated by nonlinear response \cite{gomez12,otsuki14,coulais14,boschan16,boschan17,srivastava17}, and both the yield stress \cite{olsson07,heussinger09,tighe10c} and the range of strains that are free of microscopic contact breaking events \cite{combe00,schreck11,vandeen14,vandeen16,boschan16,lerner13} all vanish.

To fill in the gap between zero and infinite strain amplitudes and shed light on the rheological scenarios and microscopic mechanisms of yielding, we perform simulations of packings under oscillatory shear at varying strain amplitude $\gamma_0$, pressure $P$, and number of particles $N$, focusing on the low frequency regime. 
Macroscopically, we identify not one but {\em two} distinct crossover strain scales. We find that linear response gives way to softening above a strain scale $\gamma_s$, after which the material's response reaches a new plateau before the material yields at a scale $\gamma_y$.  
{Unlike prior observations of two-step yielding\cite{pham06,koumakis11,segovia12}, this scenario does not require interparticle attraction.}
Near jamming softening and yielding scales
differ in their pressure dependence: while $\gamma_y$ is essentially constant, the softening strain $\gamma_s$ vanishes linearly with $P$.  Far above jamming the scales merge, suggesting the existence of a characteristic pressure that distinguishes jamming-like and dense systems. The relation between these rheological phenomena and the microscopic particle motion is complex. Whereas yielding is associated with the onset of diffusive particle motion, the relation between softening and microscopic
rearrangements is far more subtle. While rearrangements destroy strict reversibility in the particles' trajectories, bulk properties such as the storage and loss moduli remain linear long after the first rearrangements - leading to a regime where particles exhibit irregular, trapped motion but the  bulk response appears linear.  Bulk softening only becomes apparent after an extensive amount of contact breaking events; the sum of many of these singular events leads to non-trivial power law scaling of the elastic modulus with strain amplitude.

These findings provide a fresh perspective on the physics of yielding, evidence a characteristic pressure scale that distinguishes jamming-dominated systems (such as granular media and wet foams)
and dense systems (such as dry foams), and suggest that these two qualitatively different classes of soft glassy materials can be distinguished by their experimentally accessible rheological fingerprint.


\section*{Macroscopic rheology: Softening and Yielding}

We use oscillatory rheology to show that there are three distinct flow regimes, which we refer to as the linear response, softened, and yielding regimes.
A time-varying shear strain $\gamma = \gamma_0 \sin{(\omega t)}$ is applied to soft sphere packings
consisting of $N$ particles that are at pressure $P$ when unsheared,
and the resulting shear stress $\sigma$ is measured (see Appendix A).
In the linear response regime, $\sigma$ is proportional to $\gamma_0$,  and the in-phase and out-of-phase components of $\sigma/\gamma_0$ are called the storage and loss moduli $G'$ and $G''$, respectively; they characterize elastic stiffness and viscous damping. For finite strain amplitudes, the stress response ceases to be purely sinusoidal (see Appendix C), and the first harmonic terms in the Fourier expansion of $\sigma$ are used to define the storage and loss moduli, which now depend on $\gamma_0$. We focus on data taken at the low frequency $\omega = 10^{-3}$ in units constructed from the microscopic stiffness and damping coefficients; in Appendix D we demonstrate that our conclusions are unchanged at the higher frequency $10^{-2}$. Moduli are reported for data recorded after the passage of an initial transient, which can be identified from the cycle-to-cycle diffusion statistics (see below).

\begin{figure}[t!]
\begin{center}
\includegraphics[width=\linewidth,clip,trim=2.1cm 11cm 6.7cm 2.5cm]
{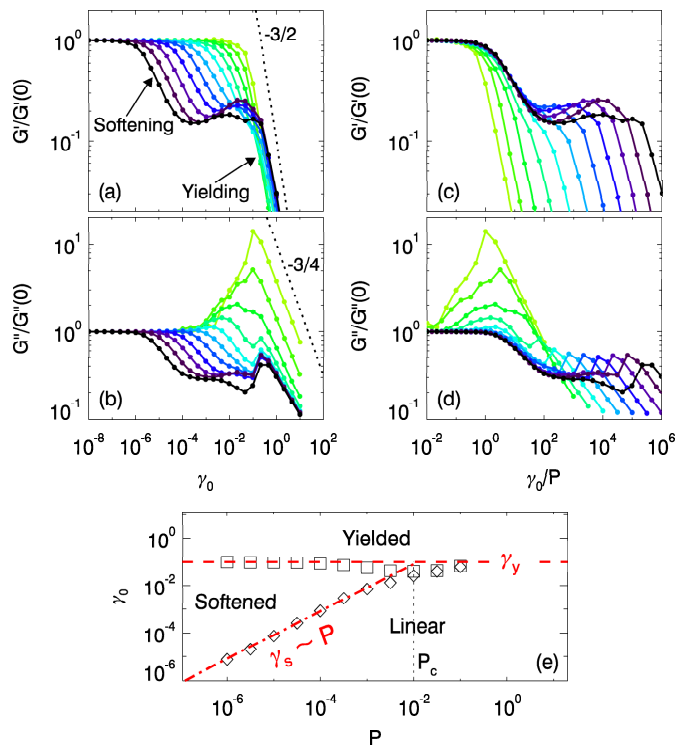}
\end{center}
\label{fig1}
\caption{Softening and yielding for  $N=1024$, $\omega=10^{-3}$ and $P$
ranging from
$10^{-6}$ (black) to $10^{-1}$ (light green) in two steps per decade. Each data point corresponds to an ensemble average
of at least 25 packings at fixed $P$ and $\gamma$. For each packing and value of $P$, $G'$ and $G''$ reach a clear plateau value for small $\gamma$ that we use to define the linear response values $G'(0)$ and $G''(0)$ (For the variation of the linear response values with $P$ and $\omega$, see Appendix B). To focus on the variation of $G'$ and $G''$ with strain amplitude, we plot
the rescaled elastic modulus $G'/G'(0)$ and rescaled loss modulus $G''/G''(0)$ as function of the strain amplitude $\gamma_0$.
(a-b) The elastic and loss moduli have a clear plateau at low strain amplitudes,
before showing softening at a pressure dependent strain scale $\gamma_s$, and yielding
at a larger strain scale $\gamma_y$. Dashed lines indicate power law decay of the moduli for large strains with slope $-3/2$ and $-3/4$ respectively.
(c-d) The softening transitions in $G'$ and $G''$ collapse when
plotted as function of the rescaled strain amplitude $\gamma_0/P$. (e) Linear, softened and yielded regimes as function of the control parameters $P$ and $\gamma_0$. Squares and diamonds
indicate yielding and softening obtained from the data for $G'/G'(0)$ shown in panel (a) (see Appendix A). The dashed lines are guides to the eye, and indicate that $\gamma_y \approx 10^{-1}$, $\gamma_s \approx 10^{-1} \times P$, leading to a crossover pressure scale $P_c \approx 10^{-2}$.
}
\end{figure}

Our simulations uncover surprisingly rich rheological scenarios, as shown in
Figs.~1a-b, which display the variation of the
rescaled storage modulus $G'(\gamma_0, P)/G'(0,P)$ and the loss modulus $G''(\gamma_0, P)/G''(0,P)$ as a function of strain - in Appendix B we demonstrate that the linear response quantities scale with pressure and frequency consistent with earlier predictions \cite{tighe11}.
Our data evidences three different rheological regimes.
{\em (i) Linear response:} For each pressure there is a finite strain range for which the moduli are essentially constant and equal to $G'(0)$ and $G''(0)$, indicative of linear response.
{\em (ii) Softening:} Surprisingly, for low pressures and
intermediate strains, both moduli fall below their linear response values but
then reach new plateau values.
We call this softening, to distinguish it from yielding, and associate a strain scale $\gamma_s$ with its onset.
{\em (iii) Yielding:} At sufficiently large strain amplitudes, the elastic modulus rapidly decays, while the loss modulus peaks. We  refer to this as yielding, with an associated yield strain $\gamma_y$ that does not strongly vary with $P$.
For large strains the storage and loss moduli fall off rapidly as
$1/\gamma_0^{\nu'}$ and $1/\gamma_0^{\nu''}$, respectively, with $\nu' \approx 1.5$ and $\nu'' \approx \nu'/2$ \cite{wyss07,datta11}.

Softening and yielding are distinct phenomena, each with their unique rheological signature and
pressure dependence. First, the loss modulus goes down at softening, and up at yielding,
signaling a qualitative difference between these two phenomena. Second,
systems at lower pressures clearly soften at smaller strain amplitude, whereas the yielding strain appears pressure independent. To characterize the pressure dependence of $\gamma_s$, we replot the data of Fig.~1a-b as a function of the rescaled strain $\gamma_0/P$ in Figs.~1c-d.  We find excellent collapse in the linear response and softening regimes of both the storage and the loss moduli for $P$ up to $10^{-2}$. Recently, several works have presented conflicting evidence for the scaling of the softening transition with pressure \cite{otsuki14,nakayama16,boschan16,goodrich16}, and
both $\gamma_s \sim P$ and $\gamma_s \sim P^{3/4}$ have been suggested - our data  is inconsistent with an exponent $3/4$.

We summarize our findings of a pressure-dependent softening transition, and a pressure-independent yielding transition in Fig.~1e. Here the red dot-dashed line indicates the softening crossover at $\gamma_s \sim P$, the red dashed line indicates the yielding crossover at $\gamma_y$, and these lines meet at a characteristic pressure $P_c$.
We note that many scaling laws near jamming break down when $P$ becomes of order $10^{-2}$ \cite{vanhecke10}, and consistent with this our data suggests that  $\gamma_y \approx 10^{-1}$ and $P_c \approx 10^{-2}$. Whereas such breakdown of scaling can be expected sufficiently far away from any critical point, we suggest that our data shows that there is a clear crossover pressure scale which separates the near-jamming and dense regimes; in the latter the physics is essentially pressure independent, and no longer controlled by the critical jamming point. Moreover, oscillatory rheology provides a specific experimental protocol to test which asymptotic regime is relevant for a given system: the jamming phenomenology is important when softening and yielding can be distinguished.


\section*{Reversible, trapped and diffusive dynamics}

To shed light on the  microscopic signatures of the rheological softening and yielding transitions, we now probe the microscopic particle trajectories. {Recent years have seen considerable effort directed towards understanding the transition from reversible to irreversible particle trajectories under cyclic driving\cite{regev13,fiocco13,keim14,regev15,priezjev16,kawasaki16,leishangthem17}. With few exceptions\cite{knowlton14}, this work has been restricted to systems far from jamming, where the two step softening/yielding scenario identified here is absent. In that case the onset of irreversibility is found to correlate with macroscopic yielding.} Here we characterize particle trajectories by stroboscopically sampling all particles at zero strain as a function of cycle number $n$.
From these stroboscopic trajectories we compute
the cycle-to-cycle squared displacements $\Delta s_1^2$, and
the cumulative squared displacements $\Delta s^2$ (see Appendix A).

\begin{figure}[tb]
\begin{center}
\includegraphics[clip,width=\linewidth,clip,trim=1.5cm 10.5cm 1.4cm 2.5cm]
{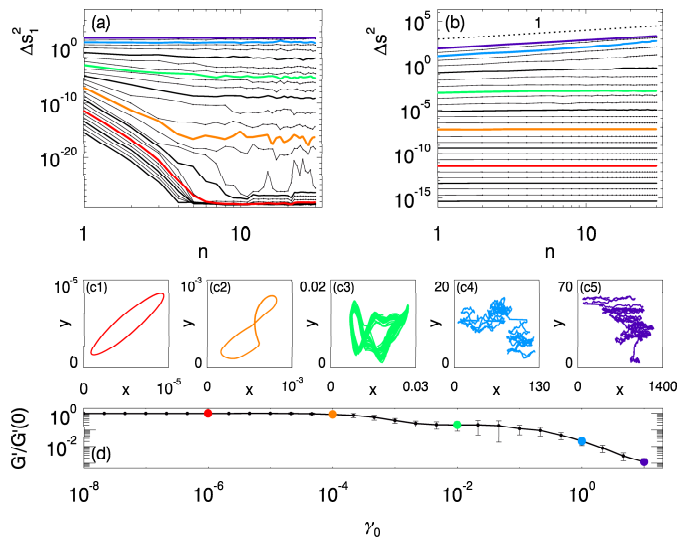}
\end{center}
\label{fig:trajectories}
\caption{Cycle-to-cycle displacements, cumulative displacements and particle trajectories for a wide range of strain amplitudes for $P=10^{-4}$, $N=128$, $\omega=10^{-3}$, and
$\gamma_0$ ranging from $10^{-8}$ to $10$   in three steps per decade.
We highlight datasets for $\gamma_0=10^{-6}$ (red),  $\gamma_0=10^{-4}$ (orange), $\gamma_0=10^{-2}$ (green), $\gamma_0=10^{0}$ (blue) and $\gamma_0=10^{1}$ (purple).
(a) Median cycle-to-cyle second moment $\Delta s_1^2$ as function of cycle number $n$ for an ensemble of 33 independent runs. $\Delta s_1^2$ rapidly decreases until it hits the noise floor for small $\gamma_0$, $\Delta s_1^2$ decays to a finite plateau for intermediate $\gamma_0$,  and $\Delta s_1^2$ is essentially constant for large $\gamma_0$.
(b) Corresponding median second moment $\Delta s^2$ as function of cycle number $n$. For small and intermediate $\gamma_0$, $\Delta s^2$ is essentially constant, dominated by the transient in early shear cycles, while for large $\gamma_0$, $\Delta s^2$ grows linear with $n$ evidencing diffusive behavior. Dotted line has slope 1.
(c1-5) Five representative particle trajectories (after a transient has been removed), for $\gamma_0=10^{-6}$ (c1, red), $\gamma_0=10^{-4}$ (c2, orange),
$\gamma_0=10^{-2}$ (c3, green), $\gamma_0=10^{0}$ (c4, blue) and $\gamma_0=10^{1}$ (c5, purple). (d) For comparison, we show $G'(\gamma_0)/G'(0)$ for $P=10^{-4}$, $N=128$ and $\omega=10^{-3}$.}
\end{figure}

Our data evidences three different dynamical regimes.
{\em (i): Reversible dynamics:} Particle trajectories at sufficiently small strain amplitudes are reversible: $\Delta s_1^2$ decays to the noise floor after an initial transient, and particles trace out ellipses in space consistent with a strictly linear response (Fig.~2c1).
{\em (ii): Trapped dynamics:}
At intermediate strains the particle trajectories are trapped: both $\Delta s_1^2$ and $\Delta s^2$ reach a finite plateau at large $n$. The particle trajectories do not form closed orbits, but remain bounded (Fig.~2c2 and Fig.~2c3).
{\em (iii): Diffusive dynamics:} For high strain amplitudes, $\Delta s^2$ grows linearly with $n$ and the
particle motion becomes diffusive  (Fig.~2c4 and 2c5).

We have verified (via the total harmonic distortion, see below) that in the reversible regime no contact changes take place during the oscillatory driving, and that the transition from reversible to trapped dynamics occurs when contacts are broken and created during the strain oscillations. Trapped dynamics is reminiscent of caging in hard sphere glasses, although there is no ballistic motion at short times and particles generally remain in contact with multiple neighbors. We note that the loss of microscopic reversibility after a single contact change suggests that the transition from reversible to caged dynamics is a finite size effect, as the strain needed to change one contact is $O(1/N)$ in large systems (verified below). Directly obtaining the characteristic strain where the first contact changes
from oscillatory rheology is numerically prohibitive, as it entails scans over  $N$, $P$ and $\gamma_0$. However recent simulations of quasistatically sheared soft spheres determined the typical strain scale $\gamma_{cc}$ at which the first contact change occurs and found that it obeys finite size scaling in the linear response regime \cite{vandeen14,vandeen16,boschan16}:
\begin{equation}
\gamma_{cc} \sim  \left \lbrace
\begin{array}{cl}
P & 			P \ll 1/N^2 \\
P^{1/2}/N &	P \gg 1/N^2 \,.
\end{array}
\right.
\label{eqn:gammacc}
\end{equation}
Our data for contact changes under oscillatory drive is consistent with this scaling, and we thus conjecture that the same scaling governs  the strain scale where the transition from reversible to caged dynamics takes place.

The pressure dependence of the transition to diffusive dynamics can be deduced from the behavior
of $\Delta s_1^2$ and $\Delta s^2$ as function of $n$; we find that the transition is essentially pressure-independent. In Fig.~3a
we plot the large $n$ plateau of $\Delta s_1^2$ as a function of $\gamma_0$ for varying pressures.
While  the initial growth of $\Delta s_1^2$, associated with contact changes, depends on $P$ -- consistent with Eq.~(1) and observations that contact breaking near jamming depends on $P$ \cite{schreck11,lerner13,vandeen14,vandeen16,boschan16} --
the asymptotic value of $\Delta s_1^2$ becomes independent of $P$ for large $\gamma_0$.
This increase signals a pressure-independent transition to diffusive motion, as is further evidenced by inspecting  our data for $\Delta s^2$ for all pressures, and fitting our data for $\Delta s^2$ as $n^\alpha$. As Fig.~3b shows, the scaling exponent $\alpha$ sharply increases with  $\gamma$ and reaches a diffusive
($\alpha=1$) regime for large strains in an essentially pressure-independent manner.

We summarize our picture for the microscopic behaviors in Fig.~3c, and now discuss the relation with the rheological behaviors shown in Fig.~1e.
Our data fully supports identifying the transition to diffusive motion with rheological yielding - both are pressure-independent and their characteristic strains are close. The correspondence between the onset of diffusion and yielding is consistent with recent experimental and numerical findings that focus on concentrated emulsions \cite{knowlton14,kawasaki16}.
This link is reminiscent of the Lindemann melting criterion: once the relative particle motions reach a significant fraction of their separation at rest, structural information is lost, particles can freely diffuse, and macroscopic rigidity vanishes. A similar correspondence between diffusion and yielding also occurs in equilibrium systems, where the fluctuation-dissipation relation establishes the relation rigorously.

Surprisingly, the micro transition from reversible to trapped dynamics and the macroscopic softening are {\em not} directly linked. In fact, the transition from reversible to trapped dynamics is a finite-size artefact. This follows from the scaling of
$\gamma_{cc}$ with $N$, which dictates that in the limit of large $N$, the onset strain for trapped dynamics is vanishingly small so that the reversible regime vanishes.
Hence, for large systems, the transition from reversible to trapped dynamics is irrelevant, which is reasonable as strict microscopic reversibility should be absent in the thermodynamic limit.
As the critical strain for softening is independent of $N$, for large systems there is a wide parameter range where the microscopic dynamics is caged, but the rheology is still effectively linear.

\begin{figure}[tb]
\begin{center}
\includegraphics[clip,width=1\linewidth,clip,trim=1.1cm 11cm 3.6cm 2.5cm]
{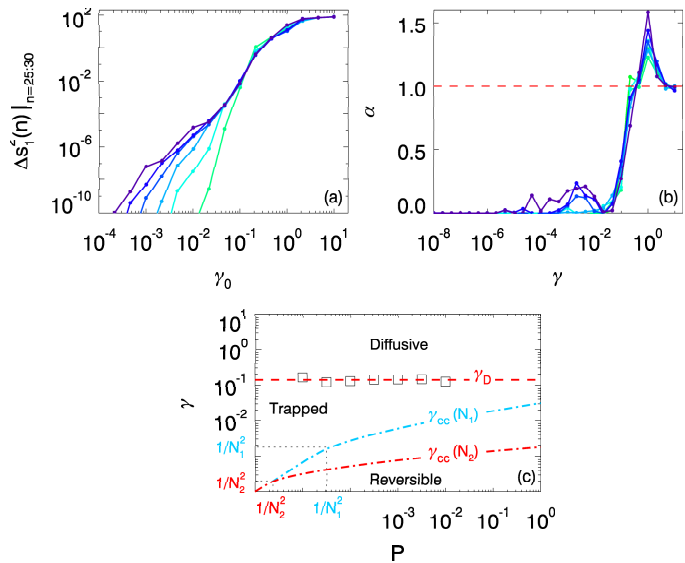}
\end{center}
\caption{Onset of diffusive motion for
$N=128, \omega =10^{-3}$, and pressures from $10^{-5}$ (purple) to $3.2 \times 10^{-3}$ (blue) in two steps per decade.
(a) Plateau values for the median cycle-to-cycle second moment, $\Delta s_1^2$, averaged over cycle $n=25-30$. The characteristic strain of the initial rise of the plateau values of $\Delta s_1^2$ increases with pressure, but the final rise  is independent from $P$. (b) We estimate the onset of diffusive motion by fitting our data for $\Delta s^2$ as $n^\alpha$ (see Appendix A), and plotting the scaling $\alpha$ exponent as function of $\gamma$; $\alpha=1$ corresponds to diffusion.
(c) {Proposed state diagram indicating reversible, trapped and diffusive regimes as function of the control parameters $P$, $\gamma_0$ and $N$. Symbols correspond to the onset of diffusive motion defined
as the strain where $\alpha$ crosses 1; red dashed line is a guide to the eye at $
\gamma=0.14$.
The dot-dashed lines indicate the prediction from Eq.~(1) for the transition from reversible to caged dynamics;  here $N_2>N_1$.
The regime $P< 1/N^2$ and $\gamma_0 < 1/N^2$ indicates the finite size regime where the
first contact change arises at $\gamma_{cc} \approx P$.
\cite{goodrich12,dagois-bohy12,goodrich14}.
Outside the finite scaling regime, $\gamma_{cc} \sim \sqrt{P}/N$ and thus vanishes for large $N$; hence the reversible regime disappears in the thermodynamic limit, where only the trapped and diffusive regimes play a role.}
}
\label{fig:RMSDpressure}
\end{figure}


\section*{Trapped motion and softening near jamming}

In the remainder of this article, we will disentangle the relation between contact breaking and
softening. Both are pressure-dependent, and both have characteristic strains that vanish when $P \rightarrow 0$ --- they are thus connected to the jamming transition. The picture that will emerge
is that contact changes have an ${\cal O}(1/N)$ effect on $G'$, and occur at strains of order $1/N$.
Significant softening occurs after an extensive number of contact change events have happened, and
this leads to a well-defined thermodynamic limit. Finally, the effect on the elastic modulus of contact breaks is cumulative, so that $G'$ decays linearly with $|\gamma_0|$, thus signifying non-analytic behavior.

{We first note that the contact change strain provides a way to rationalize the linear dependence of the softening strain on $P$ if we postulate that contact changes are necessary for softening, so $\gamma_s \ge \gamma_{cc}$. This is plausible because the shear response of packings can be mapped directly onto a spring network for infinitesimal strains. However networks {\em stiffen} rather than soften at finite strains\cite{wyart08}, which involve contact changes in packings but not in networks. We expect the lower bound on the softening strain will saturate for marginal packings that would unjam with the loss of just one contact. This marginal state is reached when $P \sim 1/N^2$\cite{tighe11b,goodrich12,goodrich14}. If we make the ansatz $\gamma_s \sim P^\nu$, the value of the exponent $\nu$ is then determined by requiring $\gamma_s \sim \gamma_{cc}$ when $P \sim 1/N^2$. The result is $\nu = 1$, consistent with the observed scaling of $\gamma_s$.

Returning to numerics,} the link between contact changes and mechanical softening is illustrated in Fig.~4a, which focuses on a single packing. We detect contact changes via the sharp increase of anharmonic behavior of the time dependent stress signal, caused by abrupt changes in the contact stiffness when harmonic contacts open or close, and quantified by the total harmonic distortion $THD:=\Sigma_{i>1} |\sigma_i^2| / |\sigma_1^2|$, where we have decomposed the stress signal in a Fourier series
with coefficients $\sigma_1,\sigma_2,\dots$. At a characteristic strain $\gamma^*$, Fig.~4a
shows a sharp increase in the $THD$ over
several orders of magnitude, accompanied by a  small decrease in $G'$. This corroborates our expectation that softening can only arise due to contact changes.
Moreover, most of the softening of the storage modulus happens subsequent to, not at, the first contact break.

To uncover the link between contact changes and softening, we now explore the role of
system size, and simulate oscillatory rheology for system sizes varying from $N = 32$ up to $N = 2048$
at fixed pressure $P=10^{-3}$, while varying the driving amplitude $\gamma_0$ over many decades. We found that $G'(\gamma_0)$, in particular for small systems, exhibits significant fluctuations, which requires a careful procedure to obtain meaningful averages. For each packing, we therefore define $\gamma_{\Delta}$ as the smallest strain
where the deviation in $G'$ from linear response, $1 - G'(\gamma)/G'(0)$ reaches a value $\Delta$, and then take ensemble averages to obtain $\gamma_{\Delta}(N,\Delta)$.

In Fig.~4b we compare the mean values of $\gamma^*$ (closed symbols) - obtained from detecting jumps in the THD - and $\gamma_\Delta$ for a range of $\Delta$ and $N$ (open symbols).
First, our data shows that $\gamma^*$ indeed decreases with system size, consistent with the $1/N$ of Eqs.~\ref{eqn:gammacc}.
Second, for very small values of $\Delta$, the data for
$\gamma_\Delta$ closely approach $\gamma^*$, consistent with the picture that any appreciable softening only arises after contact changes accumulate.
Third, for large $N$ or large $\Delta$, $\gamma_{\Delta}$ becomes independent of
$N$, evidencing a well defined continuum limit, where softening is due to an extensive number of contact changes.

As shown in Fig.~4c, all data for $\gamma_{\Delta}$ can be collapsed on a master curve by plotting
$\gamma_{\Delta}/\Delta$ as function of $N \Delta$ - data taken at different pressures shows the same trends (See Appendix D).
This scaling collapse shows, first, that the typical effect of a single contact change on $G'$ is ${\cal O} (1/N)$.
Second, the behavior for $N \Delta \ll  1 $ is consistent with $\gamma_{\Delta}/\Delta \sim (N \Delta)^{-1}$, or $\gamma_{\Delta} \sim 1/N$: hence, for small
$\Delta \ll 1/N$, $\gamma_{\Delta}$ scales as, and is close to, $\gamma^*$.
Third, the plateau for $N \Delta \gg 1$ confirms the existence of a well defined continuum limit, where $\gamma_{\Delta}$, and hence the softening behavior, becomes independent of $N$ and $\gamma_{\Delta}$ is linear in $\Delta$.
We conclude that softening first sets in once contacts start to break, only becomes significant when many contacts are changing, and is independent of system size for large $N$.

\begin{figure}[tb]
\begin{center}
\includegraphics[clip,width=\linewidth,trim=2.3cm 11.3cm 3.2cm 2.7cm]
{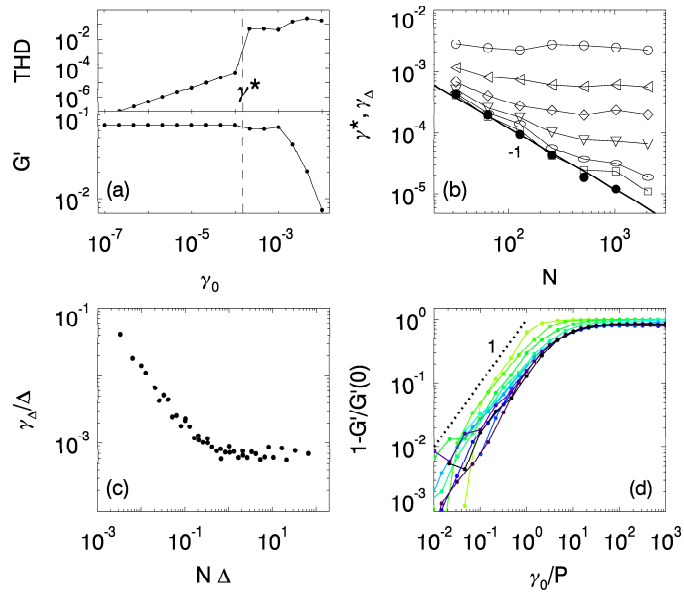}
\end{center}
\caption{(a) Total Harmonic Distortion and $G'$ as function of $\gamma_0$ for a single packing at $P=10^{-3}$, $N=128$ and $\omega=10^{-3}$.
(b) Characteristic strain scales for contact changes and softening, for
$P = 10^{-3}$, $\omega=10^{-3}$ and  $N$ ranging from 32 to 2048. The mean first contact change strain
$\gamma^*$ is determined from the jump in THD (filled circles), while $\gamma_\Delta$ for varying $\Delta$ is extracted from the elastic shear modulus for a range of $\Delta$:
$\Delta=10^{-3} (\Box)$,  $\Delta=3.2\times 10^{-3} $ (oval),
$\Delta=10^{-2} (\nabla)$,  $\Delta=3.2\times 10^{-2} (\Diamond)$,
$\Delta=10^{-1} (\lhd)$ and  $\Delta=3.2\times 10^{-1} (\circ)$.
(c) Scaling collapse for $\gamma_{\Delta}$ (same data as panel (b)).
(d) The deviation from the plateau value of the elastic modulus, $1-G'/G'(0)$, grows linearly with strain $(N=1024,\omega=10^{-3})$, $P$ ranging from $10^{-6}$ (black) to 0.1 (light green) in two steps per decade. For low $P$,
the data collapses when plotted as function of $\gamma_0/P$}
\end{figure}

A final striking consequence of the softening being caused by the accumulation of independent
contact changes is that the functional form of $G'(\gamma_0)/G'(0)$ is non-analytical.
Our picture, backed up by the various scaling collapses, suggests that $G'(\gamma_0)/G'(0)$ should decrease linearly with the strain, or equivalently, that $\Delta = 1-G'/G'(0)$ grows linearly with $\gamma$. In Fig.~4d we show data for $\Delta$ for large systems and a range of pressures, which confirms this linear deviation of the plateau value of $G'$ with strain.
As $\Delta(\gamma_0)$ needs to be an even function due to symmetry,  this implies non-analytic behavior where $\Delta \sim |\gamma_0|$. We note that similar non-analytic behavior has also been observed for
strain stiffening in random spring networks at the rigidity transition  \cite{wyart08} -- consistent with the usual association of non-analyticity with a phase transition. In contrast, the non-analytic behavior found here occurs at a finite distance from the jamming point, and we suggest that it is inherited from the purely repulsive contact forces between particles, which are themselves non-analytic at the point of contact.


\section*{Discussion}

Simulating large amplitude oscillatory shear, we have found evidence for two qualitatively distinct yielding scenarios for soft glassy solids. In dense systems, such as highly concentrated emulsions and Lennard-Jones glasses \cite{fiocco14}, the macroscopic stress-strain response is linear up to the point of yielding, which occurs at a constant strain $\gamma_y$. In marked contrast, weakly jammed solids such as wet foams and emulsions first soften at a pressure-dependent strain $\gamma_s$, only to yield at a larger strain $\gamma_y$. The ratio of $\gamma_s$ and $\gamma_y$ determines a characteristic pressure on the order of $10^{-2}$ that marks the dividing line between these two material classes.

Particle trajectories evidence an intricate link between microscopic and macroscopic behavior.
The particle dynamics display just one clear transition that separates trapped and diffusive trajectories, and which corresponds to yielding. In contrast, softening has no sharp microscopic fingerprint, but results from the accumulation
of an extensive number of contact changes leading to non-analytic rheological curves.
Our measurements correspond well with recent experiments in emulsions \cite{knowlton14} and pastes \cite{keim13}. Softening is also observed in attractive glasses \cite{koumakis11} and granular media \cite{coulais14,otsuki17}; however the characteristic strains scale differently, likely due to attraction, friction, and/or non-harmonic contact force laws. Our data for cycle-to-cycle diffusion (Fig.~2a) are strikingly similar to data from emulsions, which also show a swift rise that sharpens with increasing $P$ \cite{knowlton14}.

So far, we have focused on the behavior at a single driving frequency ($\omega=10^{-3}$). However, our data for the characteristic strains and changes in diffusive behavior obtained at $\omega=10^{-2}$ are very close, as we detail in Appendix D. In both cases, the elastic contributions to the stress are dominant and well separated from the viscous and (yet smaller) inertial contributions. This suggests that our scenario describes a rate independent, low frequency regime.
We note that recent simulations of transient rheology find similarly rate-independent characteristic strains at these frequencies \cite{boschan16}, and that the existence of rate independent characteristic strains is to be expected above jamming, when the system is solid.

Our findings clarify the notion of linear response \cite{schreck11,goodrich14b}. In sufficiently large systems, vanishingly small strains lead to contact changes, perfectly reversible trajectories are not to be found, and linear response in the strict sense, in which the microscopic equations of motion can be linearized about  the initial condition, is violated as soon as the first contact change occurs \cite{combe00,schreck11,goodrichschreckcomment,lerner13,goodrich14b,vandeen14,vandeen16,boschan16}.
However, we have provided conclusive evidence that the breaking of
contacts does not significantly influence the macroscopic behavior \cite{goodrich14b,vandeen14,boschan16,boschan17},
leading to a well-defined effective linear response for macroscopic quantities.

\section*{Conflicts of interest}
There are no conflicts of interest to declare.

\section*{Acknowledgements}
{S.D.-B.~acknowledges funding from the Dutch physics foundation FOM. E.S.~was supported by the Hungarian National Research, Development and Innovation Office NKFIH under grant OTKA K 116036.
B.P.T. ~and M.v.H.~acknowledge financial support from the Netherlands Organization for Scientific Research (NWO).}

\section{Appendix A: Numerical Model}

We perform MD simulations of 2D packings submitted to oscillatory shear.
The initial packing configurations are so-called shear stabilized (SS) packings \cite{dagois-bohy12,goodrich14}. Unlike algorithms that relax particles within a fixed box, SS packings are equilibrated in a purely isotropic stress state; the shear modulus is guaranteed to be positive, and there are no residual shear stresses. This is crucial, because packings with residual stress subjected to oscillatory driving are prone to long transients, making the use of SS packings crucial for numerical studies of oscillatory rheology.

The shear is imposed using Lees-Edwards periodic boundary conditions, and Newton's Laws are integrated using a velocity-Verlet algorithm modified for velocity dependent forces.
Our packings are composed of $N$ soft spheres of radii linearly spaced from 1 to 1.4 and mass density $\rho=1$. The spheres interact through contact elastic repulsion (amplitude $f_{el} = k\delta$, where $\delta$ is the overlap between two particles and $k=1$), as well as viscous damping ($\vec f_v = -b \, \Delta \vec v$, where $\Delta \vec v$ is their velocity difference at contact and $b=1$).

We measure the steady state stress tensor
\begin{equation}
\sigma_{\alpha\beta} =
\frac{1}{V}\sum_{\{i,j\}}  f_{ij}^\alpha \, \ r_{ij}^\beta
+ \frac{1}{V} \sum_i m_i v_i^\alpha \, v_i ^\beta \,,
\end{equation}
that develops in response to an imposed shear strain $\gamma_{xy}(t) = \gamma_0\sin(\omega t)$. The first sum is over contacts, $\vec f_{ij}$ is the sum of elastic and viscous contact forces between particles $i$ and $j$, and $\vec r_{ij}$ points between their centers. The second sum gives the kinetic stress, where $m_i$ is mass and $\vec v_i$ is velocity. In the low-frequency parameter regimes we explore, the kinetic stress is orders of magnitude below the viscous and elastic stresses, and the latter dominates.

We determine $G^*$ via the complex ratio of the first harmonic in the stress and strain signals, carefully checking that we
reach stationarity.
As our numerical model is deterministic and well behaved, statistical error bars are set by numerical noise and generally very small. However, it is well known that ensembles of finite size jammed systems at fixed pressure exhibit a considerable spread in quantities such as the static shear modulus \cite{dagois-bohy12,goodrich12,goodrich14,vandeen16}. Our data for $G'$ and $G''$, both in linear response and at finite strain, show concomittantly strong ensemble fluctuations, and we report mean values $\left< G'(\gamma)/G'(0)\right>$ and $\left< G''(\gamma)/G''(0)\right>$.

{\em Diffusivity:}
Our local dissipation law is numerically expensive but crucial to capture the correct physics \cite{tighe10c}, and to properly resolve each oscillatory cycle, a large number of simulation steps are needed, in particular near jamming. Therefore, these
simulations are numerically expensive, and  pushing our numerical capacities, to obtain particle tracks
we have simulated 33 realizations of systems of $N=128$ particles
for each value of $P$, $\Gamma_0$ and two values of $\omega$; for $\omega=10^{-3}$ we
have simulated 30 cycles, and for  $\omega=10^{-2}$, where transients are longer-lived, 300 cycles.

For each realization
we calculate the second moment  $ds^2(n,m) = \langle (x^i_n - x^i_m)^2 + (y^i_n - y^i_m)^2 \rangle$, where $m,n$ are cycle numbers and $\langle \cdot \rangle$ is the average over the non-rattler particles. These measures are highly sensitive to rattlers and drift in the center of mass, which is a Goldstone mode. {
Before processing the position data, we first carefully identify the rattling particles at each time step and remove them from any analysis,}
\textit{then} compute the Goldstone mode and remove it from the non-rattler positions. Nevertheless,
some runs still exhibit residual drift or rattlers. As the reported second moments vary over 30 decades, drift makes the main trends less easy to observe. Therefore we focus on the median of the distributions
of $ds^2(n,m) $, in particular the cycle-to-cycle second moment $\Delta s_1^2(n):=\left[ds^2(n-1,n)\right]$ and squared displacements $\Delta s^2(n):=\left[ds^2(0,n)\right]$, where
square brackets denote the median. In Appendix D we also show results for the mean, which show more fluctuations but do not lead to a different interpretation.

\begin{figure}[t]
\begin{center}
\includegraphics[clip,width=\linewidth,clip,trim=0cm 0cm 0cm .2cm]{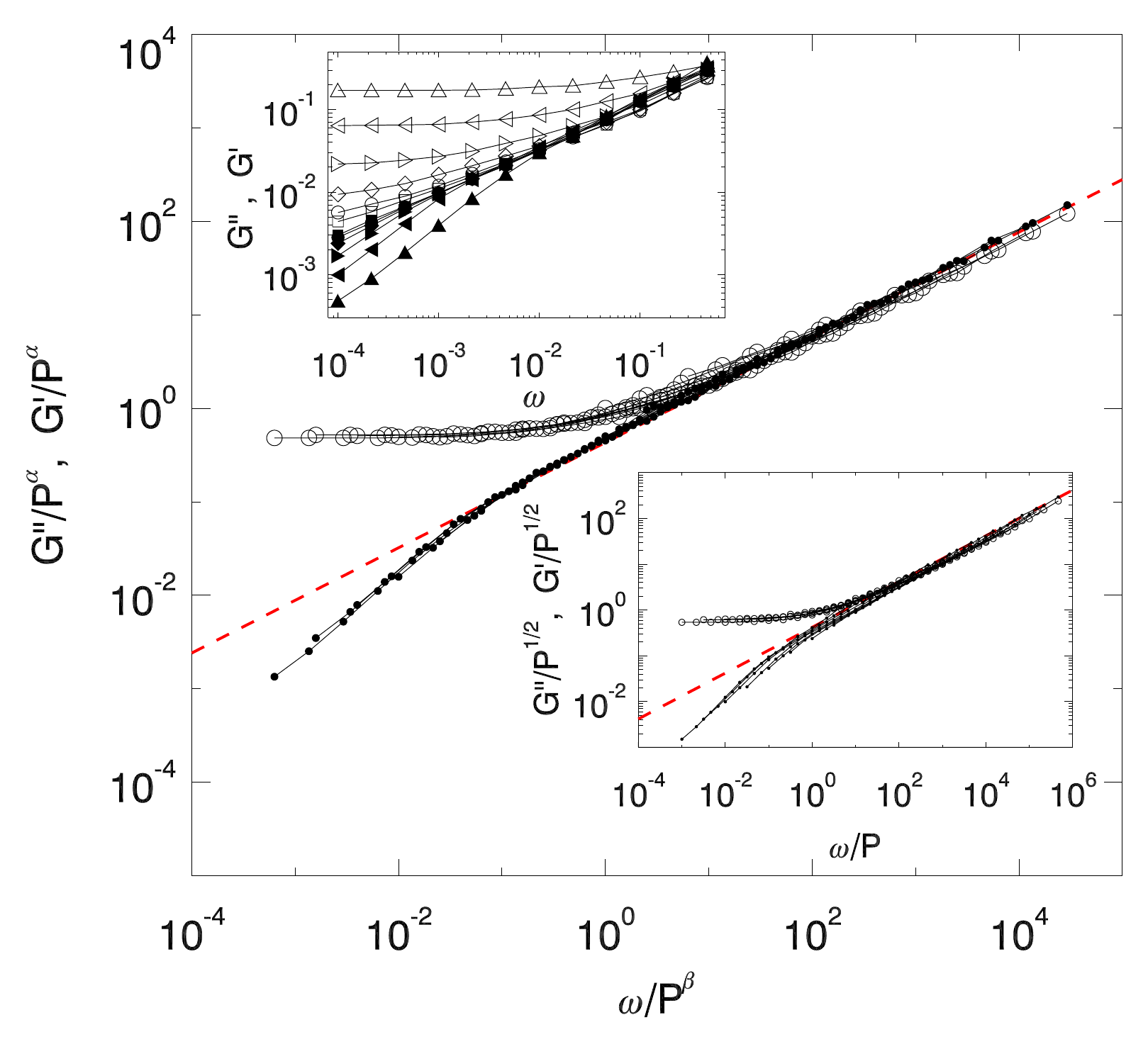}
\end{center}
\caption{Scaling collapse of $G'/P^\alpha$ and $G''/P^\alpha$ when
plotted vs $\omega/P^\beta$, where $\alpha \approx 0.45$, $\beta \approx 0.8$. The dashed line has slope $\alpha/\beta$. The strain is fixed at
$\gamma_0=10^{-10}$, $P$ ranges from $10^{-6}$ ($\Box$) to $10^{-1}$ ($\triangle$) in one step per decade, $\omega$ ranges from $10^{-4}$ to $0.46$, and $N=1024$. Left inset: Unscaled data for $G'$ (open symbols) and $G''$ (closed symbols). Right inset: Approximate scaling collapse with  $\alpha=1/2$, $\beta=1$.\label{SIfig1}}
\end{figure} 

{\em Determination of softening, yielding and diffusive onsets:}
We determine the data points for softening and yielding shown in Fig.~1e as follows. First, from our rheological data we estimate the strain where $G'/G'(0)$ dips below 0.3 and 0.03 respectively. Second,
to obtain an estimate of the critical strains that does not strongly depend on the choice of this cutoff, we assume that  $G'/G'(0) \sim \gamma^{-3/2}$, and determine the strain where $G'/G'(0)=1$, which then gives the hypothetical intersection of the plateau at low strains and power law decay at larger strains. To determine the onset of diffusive motion shown in Fig.~3c, we take data for $\Delta s^2$ as function of $n$, and perform linear fits to $\log(\Delta s^2)$ as  $\alpha \log(n)$, focusing on $3 \le n \le 30$. We have checked that such fits are close to the data. The overshoot of $\alpha$ for intermediate strains is likely due to transients and is not expected to persist for larger $n$.

\section{Appendix B:  Scaling of $G^*$} We report here our numerical results for the linear elastic and loss moduli, measured
at  $\gamma_0 =10^{-10}$. Here the stress signal is well described by a simple harmonic response of the form $\sigma_{xy}=G^* \gamma_0 \exp(i \omega t) + c.c.$. In this regime, recent theoretical arguments predict precise scaling laws for $G'$ and $G''$ as function of $P$ and $\omega$\cite{tighe11,baumgarten17b}. In Fig.~\ref{SIfig1} we show our rheological data for a range of pressures and frequencies for a system of $N=1024$ particles. Our data is in good agreement with the aforementioned scaling arguments: (1) all data collapses when plotted as function of $\omega/P$; (2) for small values of $\omega/P$, the elastic modulus $G' \sim P^{1/2}$ \cite{ohern03,vanhecke10}
and the loss modulus vanishes linearly with $\omega$ as $G'' \sim \omega/P^{1/2}$;\cite{tighe11,baumgarten17}
at large $\omega/P$, both $G'$ and $G''$ exhibit nontrivial scaling with $\sqrt{\omega}$\cite{tighe11,boschan16,baumgarten17}.

We note that we can improve on the quality of the data collapse when we plot
$G^*/P^\alpha$ vs $\omega/P^\beta$, and find the best collapse for $\beta=0.8$, $\alpha=0.45$. We do not believe that this constitutes a significant deviation from the theoretical
mean field exponents $\alpha=1/2,\beta=1$ --- even linear response calculations show  slight deviations\cite{tighe11},
and data in 2D may suffer from log corrections, as 2 is believed to be the upper critical dimension for jamming\cite{goodrich12}. 
Moreover, our value of $\beta$ is strongly influenced by the data at the lowest $\omega$, for which simulations are expensive and we only have a limited number of oscillation cycles. Furthermore, several effects limit our scaling range. First, for $P>0.1$ we see substantial deviations from scaling, as is the case for many static properties near jamming, while for $P<10^{-6}$ finite size effects start to dominate for our case of $N=1024$ particles\cite{tighe11b,dagois-bohy12,goodrich12,goodrich14,boschan16,baumgarten17b}.
Second, for our choice of microscopic parameters, inertial effects become detectable for $\omega \gtrsim 0.05$ --- if we limit our data to a smaller range of $\omega$ and $P$, the collapse becomes better but the scaling range shrinks.

\begin{figure}[t!]
\begin{center}
\includegraphics[width=1\linewidth,clip,trim=.1cm .1cm .2cm .1cm]{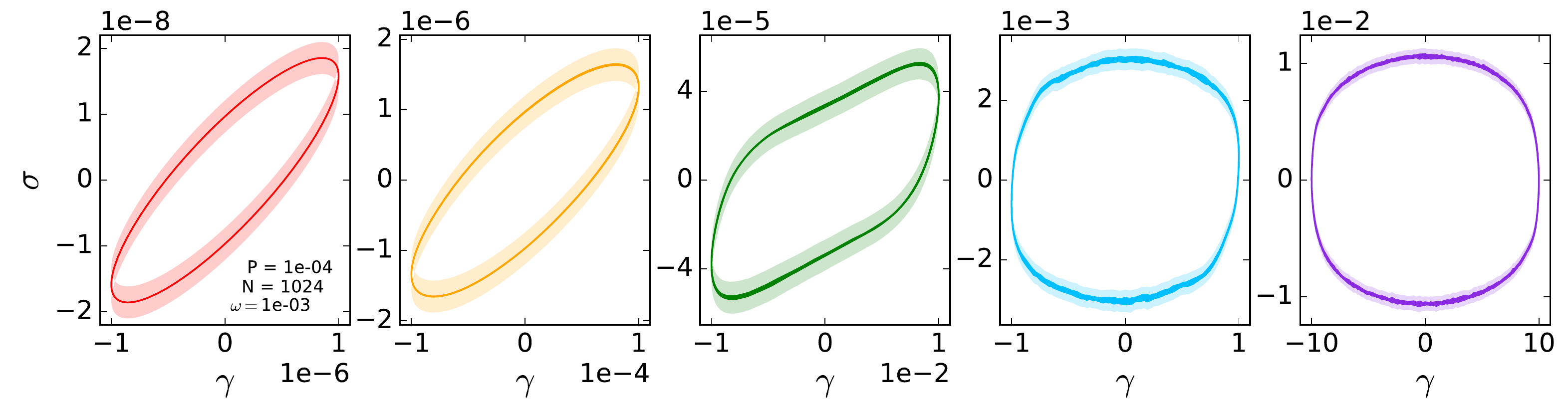}
\end{center}
\caption{Lissajous curves for $P=10^{-4}$, $N=1024$ and $\omega=10^{-3}$. Averages are shown as thick curves, and the broader, fainter band indicates the standard deviation.
}
\label{SIfigL}
\end{figure}

\section{Appendix C: Nonlinearity of response} To illustrate the nonlinearity of the large amplitude oscillatory shear response, in Fig.~\ref{SIfigL} we present Lissajous curves (stress-strain plots parameterized by time) for the same strain amplitudes presented in Fig.~2c. Anharmonic contributions are clearly visible for large strain amplitudes, but are by no means dominant, so that $G'$ and $G''$ remain meaningful.

\section{Appendix D: Robustness of results} To show that the separation of softening and yielding is robust to changes in particle number and frequency, in Figs.~7 and 8 we show our data for $G'$ and $G''$ for $N=128$ and both $\omega=10^{-3}$ and $\omega=10^{-2}$ - all features shown in Fig.~1 for $N=1024,\omega=10^{-3}$ are also present here.

\begin{figure}[t!]
\begin{center}
\includegraphics[width=0.8\linewidth,clip,trim=1.6cm 13cm 1.8cm 2.5cm]
{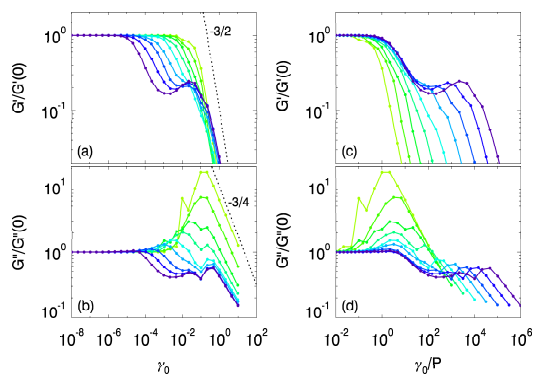}
\end{center}
\caption{Softening and yielding, characterized by the effective elastic modulus $G'$ and loss modulus $G''$ as function of strain amplitude $\gamma_0$,
for $N=128$, $\omega=10^{-3}$ and 9 values of $P$ ranging from
$10^{-5}$ (purple) to $10^{-1}$ (light green) in two steps per decade.
}
\label{SI2}
\end{figure}

\begin{figure}[t]
\begin{center}
\includegraphics[width=0.8\linewidth,clip,trim=1.6cm 13cm 1.8cm 2.5cm]
{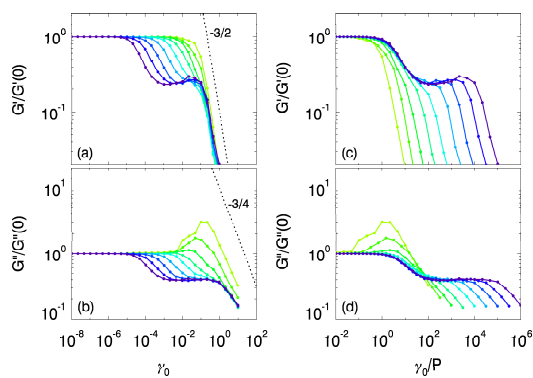}
\end{center}
\caption{Softening and yielding, characterized by the effective elastic modulus $G'$ and loss modulus $G''$ as function of strain amplitude $\gamma_0$,
for $N=128$, $\omega=10^{-2}$ and 9 values of $P$ ranging from
$10^{-5}$ (purple) to $10^{-1}$ (light green) in two steps per decade.
}
\label{SI3}
\end{figure}

{\em Diffusion:} In Fig.~9 we show examples of the median cycle-to-cycle squared displacements $\Delta s_1^2$, and
the cumulative squared displacements $\Delta s^2$ for $\omega=10^{-2}$, illustrating that
all features shown in Fig.~2 for $\omega=10^{-3}$ are also present here. We note that the jumps visible in a few datasets
are due to the large packing-to-packing fluctuations in the transition from trapped to diffusive particle dynamics.
In Figs.~10 and 11 we show the {\em mean} cycle-to-cycle squared displacements $\Delta s_1^2$ and
the cumulative squared displacements $\Delta s^2$ for $\omega=10^{-2}$ and $\omega=10^{-3}$ --- even though the mean is more sensitive to fluctuations, in particular for the smallest squared displacements (notice the large dynamical range), we stress that
all essential features are similar in mean and median data.

\begin{figure}[t]
\includegraphics[clip,width=0.98\linewidth,clip,trim=1.5cm 10.5cm 1.4cm 2.45cm]
{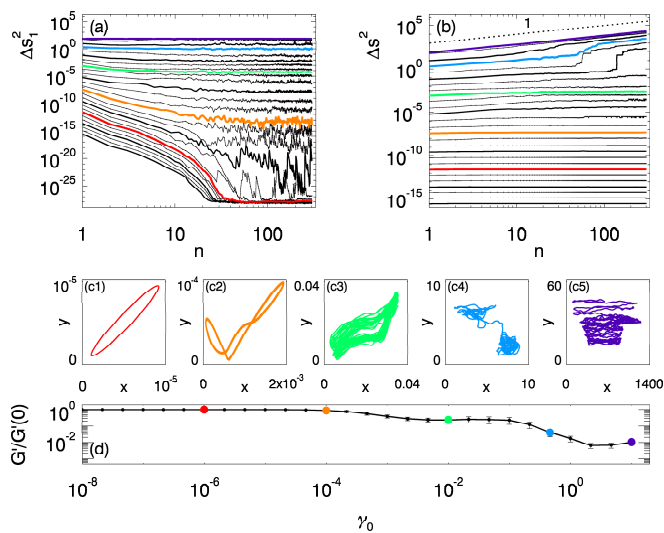}
\label{fig2}
\caption{Diffusion,  particle trajectories, and softening for a wide range of strain amplitudes, $P=10^{-4}$, $N=128$ and $\omega=10^{-2}$. In all panels, we highlight datasets for $\gamma_0=10^{-6}$ (red),  $\gamma_0=10^{-4}$ (orange), $\gamma_0=10^{-2}$ (green), $\gamma_0=0.46$ (light blue) and $\gamma_0=10^{1}$ (dark blue).
(a) Median intercycle second moment $\Delta s_1^2$ as function of cycle number $n$
for an ensemble of xxx runs,
for  $\gamma_0$ ranging from $10^{-8}$ to $10$   in three steps per decade. $\Delta s_1^2$
rapidly decreases until it hits the noise floor for small $\gamma_0$, $\Delta s_1^2$ decays to a finite plateau for
 intermediate $\gamma_0$,  and $\Delta s_1^2$ is essentially constant for large $\gamma_0$.
 (b) Corresponding median second moment $\Delta s^2$. For small and intermediate $\gamma_0$, $\Delta s^2$
 is essentially constant, dominated by the transient in early shear cycles, while for large
 $\gamma_0$, $\Delta s^2$ grows linearly with $n$ evidencing diffusive behavior (dashed line).
(c) Five representative particle trajectories (after a transient has been removed). For $\gamma_0=10^{-6}$ (red)
the trajectory is elliptical and reversible. For
 $\gamma_0=10^{-4}$ (orange), the trajectory becomes strongly nonlinear. For
 $\gamma_0=10^{-2}$ (green) the trajectory is no longer closed but remains bounded. For
 $\gamma_0=10^{0}$ (light blue), the particle motion becomes diffusive, characterized by hoping between different cages. For $\gamma_0=10^{1}$ (dark blue), the particle makes large excursions between cycles and diffuses freely.
 (d) For comparison, we show $G'(\gamma_0$) for $P=10^{-4}$, $N=128$ and $\omega=10^{-2}$. }
\end{figure}

\begin{figure}[tb]
\includegraphics[clip,width=0.98\linewidth,clip,trim=1.45cm 18.5cm 1.4cm 2.45cm]
{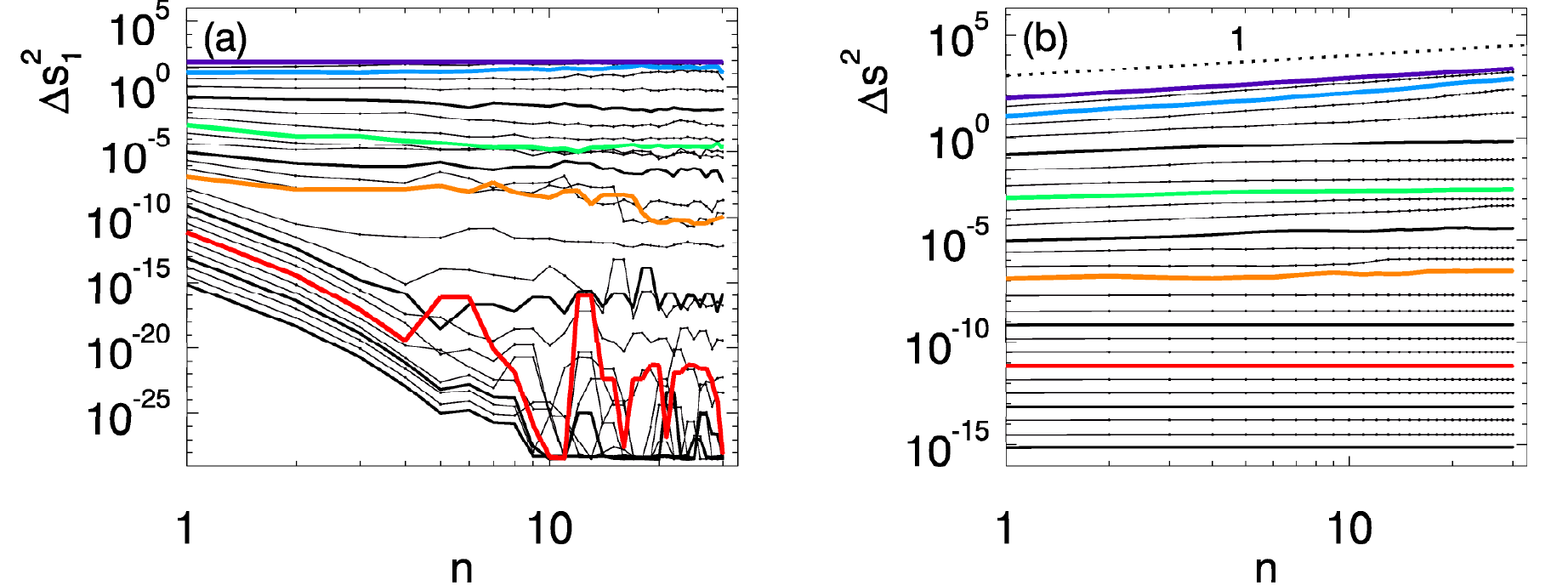}
\caption{Mean of cycle-to-cycle squared displacements $\Delta s_1^2$, and
the cumulative squared displacements $\Delta s^2$ for
 $P=10^{-4}$, $N=128$ and $\omega=10^{-3}$. In all panels, we highlight datasets for $\gamma_0=10^{-6}$ (red),  $\gamma_0=10^{-4}$ (orange), $\gamma_0=10^{-2}$ (green), $\gamma_0=10^{0}$ (light blue) and $\gamma_0=10^{1}$ (dark blue).
}
\end{figure}

\begin{figure}[tb]
\includegraphics[clip,width=0.98\linewidth,clip,trim=1.45cm 18.5cm 1.4cm 2.45cm]
{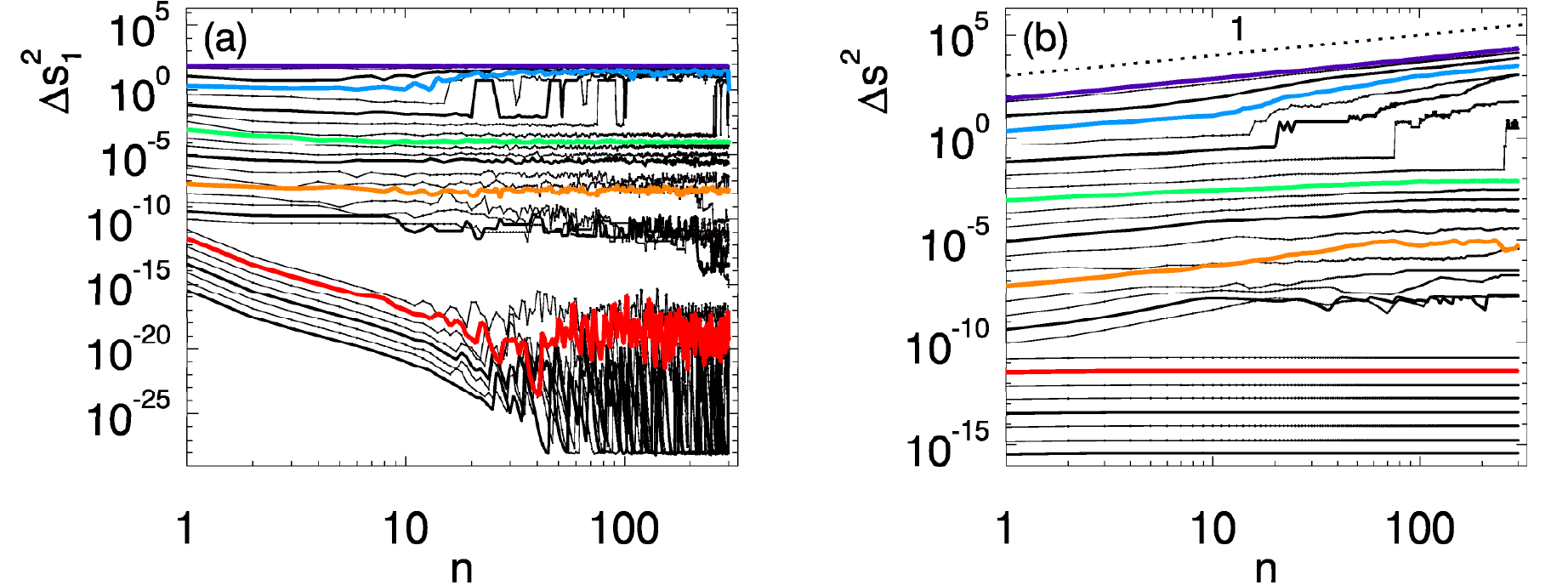}
\caption{Mean of cycle-to-cycle squared displacements $\Delta s_1^2$, and
the cumulative squared displacements $\Delta s^2$ for  $P=10^{-4}$, $N=128$ and $\omega=10^{-2}$. In all panels, we highlight datasets for $\gamma_0=10^{-6}$ (red),  $\gamma_0=10^{-4}$ (orange), $\gamma_0=10^{-2}$ (green), $\gamma_0=0.46$ (light blue) and $\gamma_0=10^{1}$ (dark blue).
 }
\end{figure}





\bibliographystyle{rsc} 

\providecommand*{\mcitethebibliography}{\thebibliography}
\csname @ifundefined\endcsname{endmcitethebibliography}
{\let\endmcitethebibliography\endthebibliography}{}

\end{document}